\documentclass[letterpaper,english,reprint,aps,pra,superscriptaddress]{revtex4-1}
\usepackage[T1]{fontenc}
\usepackage[latin9]{inputenc}
\usepackage{color}
\usepackage{bm}
\usepackage{amsmath}
\usepackage{amssymb}
\usepackage{graphicx}
\usepackage{setspace}

\makeatletter

\usepackage{bm}
\usepackage{graphicx}
\def\oper{{\mathchoice{\rm 1\mskip-4mu l}{\rm 1\mskip-4mu l}
		{\rm 1\mskip-4.5mu l}{\rm 1\mskip-5mu l}}}
\def\<{\langle}
\def\>{\rangle}
\newtheorem{Theorem}{Theorem}
\newtheorem{Proposition}{Proposition}

\usepackage{tikz}
\usepackage{pgfplots}
\usetikzlibrary{positioning}
\usetikzlibrary{decorations.markings}
\usetikzlibrary{arrows,decorations.text}
\usetikzlibrary{backgrounds,fit,decorations.pathreplacing}

\usepackage{babel}

\@ifundefined{showcaptionsetup}{}{%
	\PassOptionsToPackage{caption=false}{subfig}}
\usepackage{subfig}
\makeatother

\usepackage{babel}
\begin{document}
	
	\title{Detection power of separability criteria  based on a correlation tensor: a case study}

	\author{Gniewomir Sarbicki}
	
	\affiliation{Institute of Physics, Faculty of Physics, Astronomy and Informatics,
		Nicolaus Copernicus University, Grudziadzka 5/7, 87-100 Toru\'{n},
		Poland}
	
	\author{Giovanni Scala}
	
	\affiliation{Dipartimento Interateneo di Fisica, Università degli Studi di Bari,
		I-70126 Bari, Italy}
	
	\affiliation{INFN, Sezione di Bari, I-70125 Bari, Italy}
	
	\author{Dariusz Chru\'{s}ci\'{n}ski}
	
	\affiliation{Institute of Physics, Faculty of Physics, Astronomy and Informatics,
		Nicolaus Copernicus University, Grudziadzka 5/7, 87-100 Toru\'{n},
		Poland}
	
	\date{\today}
	\begin{abstract}
Detection power of separability criteria  based on a correlation tensor is tested within a family of generalized isotropic state in $d_1 \otimes d_2$. For $d_1 \neq d_2$ all these criteria are weaker than positive partial transposition (PPT) criterion. Interestingly, our analysis supports the  recent conjecture that a criterion based on symmetrically informationaly complete positive operator-valued measure (SIC-POVMs) is stronger than realignment criterion. 
	\end{abstract}
	
	\pacs{33.15.Ta}
	
	\keywords{detect entanglement, SIC-POVMs, realignment criterion, entanglement witnesses}
	\maketitle
	
\section{\label{sec:level1}Introduction}

Quantum entanglement defines a key feature of quantum
theory. It becomes a crucial resource for quantum information theory and  for modern quantum based 
technologies like quantum communication, quantum cryptography,
and quantum calculations \cite{HHHH,QIT}. There are several separability criteria developed during the last two decades which enable to distinguish between separable and entangled states (cf. \cite{GT,HHHH,TOPICAL}). In this paper we analyze a special family of separability criteria based on correlation tensor. The prominent example is provided be the realignment criterion or computable cross-norm (CCNR) criterion \cite{R1,R2,R3}. Apart from CCNR this class contains  de Vicente criterion {(dV)} \cite{Vicente}, separability criterion derived in \cite{Fei}  and recent criterion based on SIC POMVs (ESIC) \cite{GUHNE}. Interestingly, all these criteria were unified in \cite{GGD1}. It was shown \cite{GGD1} that the above criteria are just special examples of a 2-parameter family of criteria. In this paper we provide a comparative anaysis of these criteria testing a simple family of bipartite states in $d_1 \otimes d_2$. 

Let us fix notation: $\mathbb{C}^{\mathrm{d_{1}}}\otimes\mathbb{C}^{\mathrm{d_{2}}}$
be the Hilbert space of a bipartite system with dimension of subsystems
	$d_{1}$ and $d_{2}$ respectively. In what follows we assume that $d_2 \geq d_1$. Let $\{G_{\alpha}^{(1)}\}_{\alpha=1}^{d_{1}}$
	and $\{G_{\beta}^{(2)}\}_{\beta=1}^{d_{2}}$ be arbitrary orthonormal
	bases in $\mathcal{B}(\mathbb{C}^{\mathrm{d_{1}}})$ and $\mathcal{B}(\mathbb{C}^{\mathrm{d_{2}}})$,
	namely $\langle G_{\alpha}^{(1)}|G_{\alpha'}^{(1)}\rangle_{{\rm HS}}=\delta_{\alpha,\alpha'}$
	and the same for $G_{\beta}^{(2)}$, where $\langle X|Y\rangle_{\mathrm{HS}}={\rm Tr(X^{\dagger}Y)}$
	is the Hilbert-Schmidt inner product. One defines a correlation matrix
	\begin{equation}
	C_{\alpha\beta}=\langle G_{\alpha}^{(1)}\otimes G_{\beta}^{(2)}\rangle_{\rho}={\rm Tr\left(\rho G_{\alpha}^{(1)}\otimes G_{\beta}^{(2)}\right)}.\label{key}
	\end{equation}
Let us now restrict ourselves to orthonormal bases (referred as \textit{canonical bases}) such that $G^{(i)}_0 = \boldsymbol{1}_{\mathrm{d_{i}}}/\sqrt{d_{i}}$ (the remaining basis elements are then pairwise orthogonal hermitian traceless operators of norm one). Since now $C^{\mathrm{can}}$ stands for the correlation matrix in canonical bases. 
	%
One proves  \cite{GGD1}) the following
\begin{Theorem} \label{TH-1} If $\rho$ is separable, then
		\begin{equation}  
		\|D_{x}^{(1)}C^{{\rm can}}D_{y}^{(2)}\|_1 \leq \mathcal{N}_{x,d_{1}}\mathcal{N}_{y,d_{2}},\label{xy}
		\end{equation}
		where
		\begin{equation}
		D_{x}^{(1)}=\mathrm{diag}\left\{ x,1,\dots,1\right\} ,\qquad D_{y}^{(2)}=\mathrm{diag}\left\{ y,1,\dots,1\right\}
		\end{equation}
		and
		\begin{equation}
		\mathcal{N}_{x,d_{1}}=\sqrt{\frac{d_{1}-1+x^{2}}{d_{1}}},\,\ \ \mathcal{N}_{y,d_{2}}=\sqrt{\frac{d_{2}-1+y^{2}}{d_{2}}},\label{NANB}
		\end{equation}
		for arbitrary $x,y\geq0$. 
\end{Theorem}
Interestingly, for $(x,y)=(1,1)$ the above criterion reduces to CCNR. Moreover, for  $(x,y)=(0,0)$, $(x,y)=\sqrt{2/d_{1}},\sqrt{2/d_{2}})$, and
	 $(x,y)=(\sqrt{d_{1}+1},\sqrt{d_{2}+1})$  one recovers separability criteria developed in  \cite{Vicente}, \cite{Fei}  and \cite{GUHNE}, respectively. The last criterion \cite{GUHNE} was constructed in terms of SIC POMVs (ESIC criterion) and it was conjectured that it is stronger than original CCNR criterion. In this paper we provide a comparative analysis of CCNR, ESIC and dV criteria for a class of bipartite states being a generalization of well known isotropic states
\begin{align}
	\rho_{p} & =\frac{1-p}{d_{1}d_{2}}    \oper_{d_1} \otimes \oper_{d_2} 
	 + p |\psi^+_{d_1}\rangle \langle \psi^+_{d_1}| , \label{isotropic}
\end{align}
where $|\psi^+_{d_1}\rangle = 1/\sqrt{d_1} \sum_{i=1}^{d_1} |e_i\otimes f_i\rangle$, $|e_i\rangle$ defines orthonormal basis in $\mathbb{C}^{d_1}$ and $|f_i\rangle$ defines orthonormal set in $\mathbb{C}^{d_2}$. It is well known \cite{GT} that this state is separable if and only if it is PPT which is equivalent to
\begin{equation}\label{PPT}
  p \leq \frac{1}{d_2 + 1} . 
\end{equation}
In the paper we provide the upper bound for `$p$' implied by  condition (\ref{xy}). Moreover, we show that ESIC criterion detects more entangled state than the standard CNNR criterion. This supports the conjecture made in \cite{GUHNE} that ESIC is stronger than CCNR. Finally, we analyze so called enhanced realignment criterion \cite{ZZZ}    which states that  for separable states
\begin{equation}\label{ZZZ}
  \| \mathcal{R}(\rho_{AB})\|_1 \leq \sqrt{1- {\rm Tr} \rho_A^2}\sqrt{1-{\rm Tr}\rho_B^2} ,
\end{equation}
with $\rho_A = {\rm Tr}_B \rho_{AB}$ and $\rho_B = {\rm Tr}_A \rho_{AB}$ being marginal states. Enhanced realignment
criterion turns out to be the strongest effectively computable simplification of Correlation Matrix Criterion\cite{COV-1,COV-2,COV-3} (see also \cite{COV-U} for the unifying approach). In the recent papers \cite{GGD2} we showed that this criterion is equivalent to (\ref{xy}) for all $x,y\geq 0$. Here we show intricate relation of (\ref{ZZZ}) and (\ref{xy}) using a family of isotropic states.
\section{\label{sec:level2} $XY$-criterion for isotropic states}
The main result of our paper consists in the following
\begin{Theorem} \label{I} If an isotropic state $\rho_{p}$  is separable, then $p\leq p_{xy}$, where
			
			\begin{align}
			  p_{xy} & =\Gamma\frac{\sqrt{\left(1+\tilde{x}\right)\left(1+\tilde{y}\right)}-\sqrt{\tilde{x}\left(\left(1+\gamma\right)\tilde{y}+\gamma\tilde{x}+\gamma\right)}}{1-\gamma\tilde{x}}\label{eq:pxy}
\end{align}
with
\begin{align}
			&\tilde{x}=  \frac{x^{2}}{d_{1}-1},\qquad\tilde{y}=\frac{y^{2}}{d_{2}-1}, \\
			\gamma= & \frac{\left(d_{2}-d_{1}\right)}{d_{2}\left(d_{1}-1\right)\left(d_{1}+1\right)^{2}},\quad
			\Gamma=  \frac{d_1}{d_1^2-1} \frac{\sqrt{d_1-1}\sqrt{d_2-1}}{\sqrt{d_1d_2}}\nonumber
			\end{align}			
			for arbitrary $x,y\geq0$.
\end{Theorem}
For the proof cf. Appendix.  Note, that for $x=y$ and $d_{1}=d_{2}=d$ one finds
\begin{equation}
	p_{x,x} = \frac{1}{d+1} ,
	\end{equation}
and hence one recovers the PPT condition (\ref{PPT}). In general, however, 
\begin{equation}\label{}
  p_{x,y} > \frac{1}{d_2+1} ,
\end{equation}
and hence this criterion is weaker than PPT condition. In particular, we have the following bounds for de Vicente criterion ($x=y=0$), realignment criterion $x=y=1$, and for ESIC criterion ($x=\sqrt{d_1+1}$, $y=\sqrt{d_2+1}$):
      \begin{align}
	p_{dV} 	& = \frac{d_1}{d_1^2-1} \frac{\sqrt{d_1-1}\sqrt{d_2-1}}{\sqrt{d_1d_2}} \\
	p_{R}&=\frac{\left(d_{1}^{2}-1\right)d_{2}-\sqrt{d_{1}^{3}d_{2}-3d_{1}d_{2}+d_{2}^2+1}}{d_{2}d_{1}^{3}-2d_{1}d_{2}+1}\\
	p_{E}&=\frac{2\left(d_{1}-1\right)d_{2}-\sqrt{\frac{d_{1}^{3}d_{2}^{2}-2d_{1}d_{2}^{2}+3d_{2}^{2} +\left(d_{1}^{3}-5d_1\right)d_{2}+d_{1}+1}{d_{1}+1}}}{d_{1}^2d_2-d_{1}d_2-d_{2}+1}.\label{eq:pe}
      \end{align}
We skip the expression $p_F$ for the criterion from \cite{Fei}  ($x=\sqrt{2/d_1}$,$y=\sqrt{2/d_2}$) since it is quite complicated. Again, for $d_1=d_2$ one has
$$ p_{dV} = p_R = p_E = \frac{1}{d+1} .  $$
\begin{Proposition} For $d_1 \neq d_2$ one has
\begin{equation}\label{}
  p_E < p_R ,
\end{equation}
that is , ESIC criterion detects more entangled isotropic state that realignment.
\end{Proposition}
For the proof cf. Appendix. 
\section{Comparison with Enhanced Realignment Criterion}
\begin{figure}
	\begin{centering}
		\subfloat[$\left(\mathrm{d_{1}=d_{2}}=3\right)$]{\includegraphics[width=0.24\textwidth]{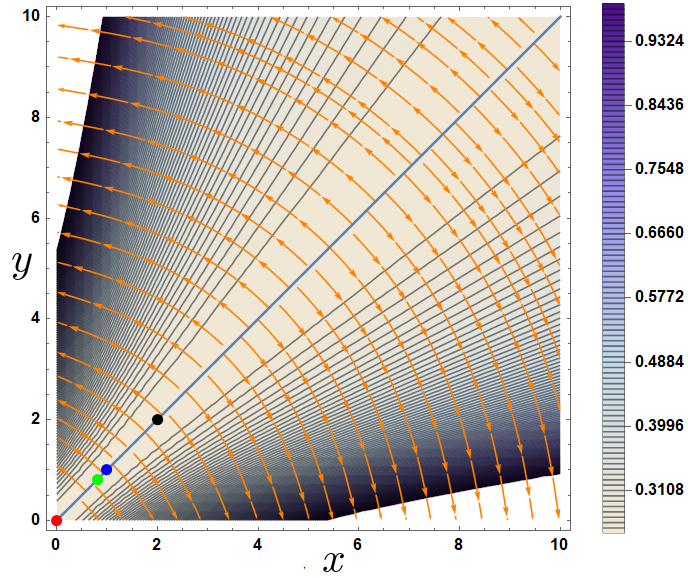}}\subfloat[$\left(\mathrm{d_{1}=d_{2}}=3\right)$]{\includegraphics[width=0.24\textwidth]{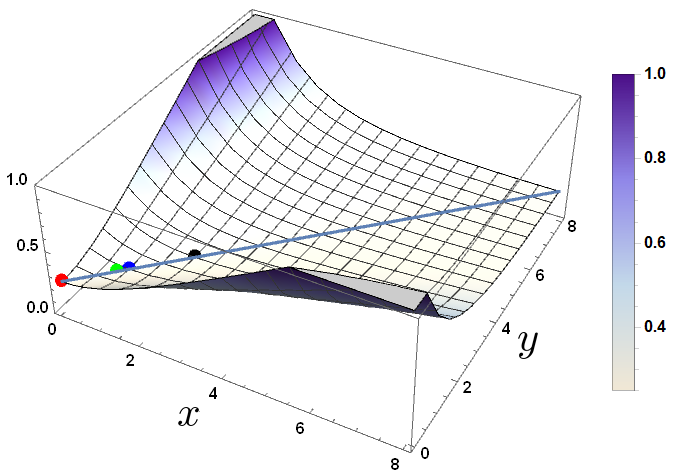}}
		\par\end{centering}
	\begin{centering}
		\subfloat[$\left(\mathrm{d_{1}=2;\:}d_{2}=20\right)$]{\includegraphics[width=0.24\textwidth]{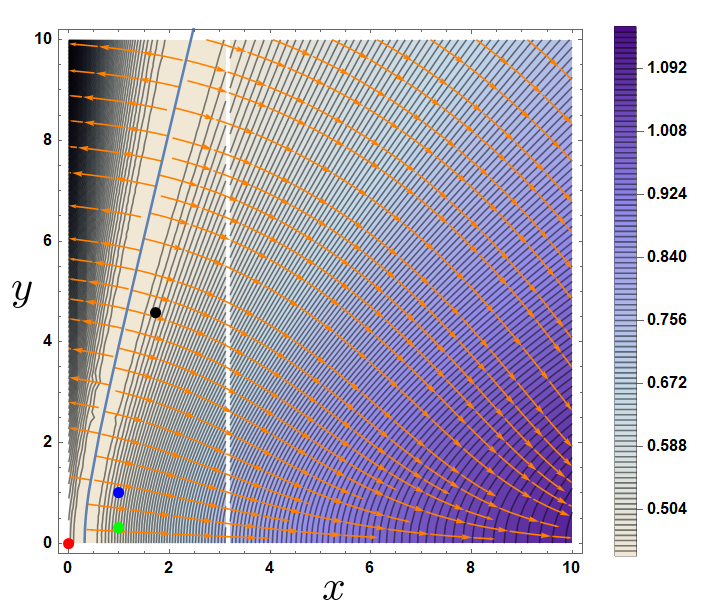}\label{c}}
		\subfloat[$\left(\mathrm{d_{1}=2;\:}d_{2}=20\right)$]{\includegraphics[width=0.235\textwidth]{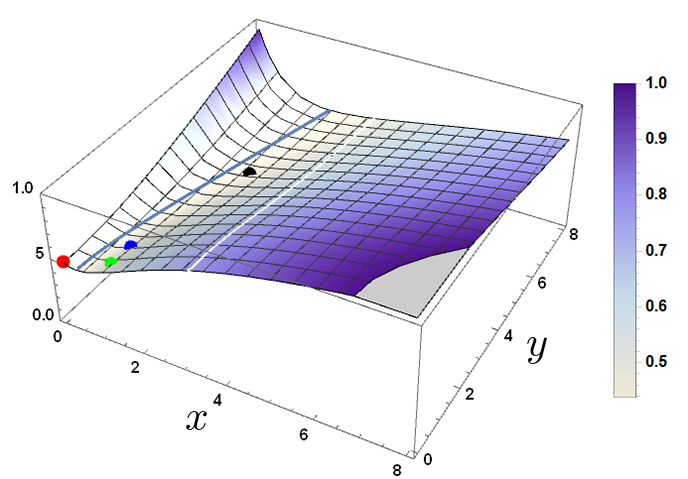}}
		\par\end{centering}
	\centering{}\caption{Contour plot with the gradient direction and plot of the threshold	function $p_{xy}$ for $\left(\mathrm{d_{1}}=\mathrm{d_{2}}=3\right)$	and $\left(\mathrm{d_{1}=2;}d_{2}=20\right)$. The colored points are in $\left(0,0\right)$ in red for Di Vincente criterion; in $\left(1,1\right)$ in blue for CCNR; in $\left(2/\sqrt{d_{1}},2/\sqrt{d_{2}}\right)$ in green for criterion in \cite{Fei}; $\left(\sqrt{d_{1}+1},\sqrt{d_{2}+1}\right)$ in black for ESIC.
		The hyperbola in blue form Eq.\eqref{eq:hyperbola}. The white line in Fig \ref{c} is for $ a=0 $ of Eq. \eqref{eq:a}.}
	\label{fig:contourplot}
\end{figure}

Separability of $\rho_p$ implies $p \leq p_{xy}$. Clearly, $p_{xy}$ depends on $(x,y)$ and hence the most efficient criterion corresponds to minimal value of $p_{xy}$. Let us calculate the minimum of the expression (\ref{eq:pxy}). One has
      \begin{equation} \label{partial_y}
	\partial_{\tilde{x}} p_{xy} = 0 \iff (1+\gamma)\tilde{y} = \tilde{x} -\gamma.
      \end{equation}
This is a necessary condition for minimum. One can check, that substituting (\ref{partial_y}) to (\ref{eq:pxy}) a constant value $\Gamma/\sqrt{1+\gamma}$ is obtained, hence we have the whole line $(1+\gamma)\tilde{y} = \tilde{x} -\gamma$ (hyperbola in $x$,$y$) of minima of $p_{xy}$. One can summarise the above observations in the following
\begin{Theorem}
	The minimum of $p_{xy}$ is attained in points of the hyperbola:
	\begin{equation} \label{eq:hyperbola}
	  \frac{x^2}{d_1-1} - (1+\gamma) \frac{y^2}{d_2-1} = \gamma
\end{equation}
and the value of the minimum reads as follows
	\begin{equation}
	  p_{\rm min} = \frac{\Gamma}{\sqrt{1+\gamma}} = \sqrt{\frac{d_2-1}{d_2(d_1^2+d_1-1)-1}}.
	\end{equation}
\end{Theorem}
The enhanced realignment criterion states that if $\rho_{p}$ is separable, 	then
	\begin{equation}
	\|\mathcal{R}(\rho_{p}-\rho_{1}\otimes\rho_{2})\|_{1}\leq\sqrt{1-{\rm Tr}\rho_{1}^{2}}\sqrt{1-{\rm Tr}\rho_{2}^{2}},\label{RR}
	\end{equation}
	where
	\begin{equation}
	\rho_{1} =\frac{\oper_{\mathrm{d_{1}}}}{d_{1}}, \qquad\rho_{2}=\frac{1-p}{d_{1}}\oper_{\mathrm{d_{2}}} + \frac{p}{d_{1}} \sum_{i=1}^{d_1} |f_i\rangle \langle f_i|
\end{equation}
are local states in the subsystems. 
\begin{Theorem} $\rho_p$ satisfies (\ref{RR}) if and only if
\begin{equation}\label{}
  p \leq p_{\rm ER} , 
\end{equation}
where
\begin{equation}
	p_{\mathrm{ER}}=\sqrt{\frac{d_{2}-1}{d_{2}\left(d_{1}^{2}+d_{1}-1\right)-1}}.\label{eq:per}
	\end{equation}
\end{Theorem}
For the proof cf. Appendix. Hence, the enhanced realignment criterion is equivalent to the to the family of $XY$-criteria minimising $p_{xy}$. Interestingly, in \cite{GGD2} we proved that this equivalence is always realized for large values of $ x $ and $ y $, but for isotropic states in Eq. \eqref{isotropic} it is enough consider $ x,y $ belonging to the hyperbola in Eq. \eqref{eq:hyperbola}. Indeed, the figure \ref{fig:contourplot} illustrates the hyperbola of minima of $p_{xy}$ (reducing to line if dimensions are equal) and four characteristic points representing the four criteria distinguished in the literature.
\section{Conclusions}
We provided a comparative analysis of the detection power of a $XY$-family of separability criteria in the case of generalized isotropic state in $d_1 \otimes d_2$. Due to the high symmetry of the isotropic state one can derive analytical formula for the separability bound $p_{xy}$ which for $d_1 \neq d_2$ is always higher than a PPT bound $p_{\mathrm{PPT}} = 1/(d_2+1)$. Interestingly, minimising over $(x,y)$ we showed that the most efficient $p_{\rm min}$ is exactly the same as the one derived in term of so called enhanced realignment criterion. Finally, it is shown that for a family of generalized isotropic states the ESIC criterion from \cite{GUHNE} detects more entangled state than original CCNR criterion. Hence, it supports a conjecture raised in \cite{GUHNE} that ESIC is stronger than CCNR. It is clear that a similar analysis can be performed for a Werner-like state \cite{Werner}
\begin{equation}\label{}
  \rho_q = \frac{1-q}{d_1 d_2 } \oper_{d_1} \otimes \oper_{d_2} + \frac{q}{d_1} \sum_{i,j=1}^{d_1} |e_i \rangle \langle e_j| \otimes |f_j \rangle \langle f_i| .
\end{equation}
As a resume of this simple analysis, which  provides a clear illustration of the detection power for a family of separability criteria based on correlation tensor, on figure is shown the general trend for large value of $ d_1 $ and $d_2$.
	\begin{figure}
	\includegraphics[width=0.48\textwidth]{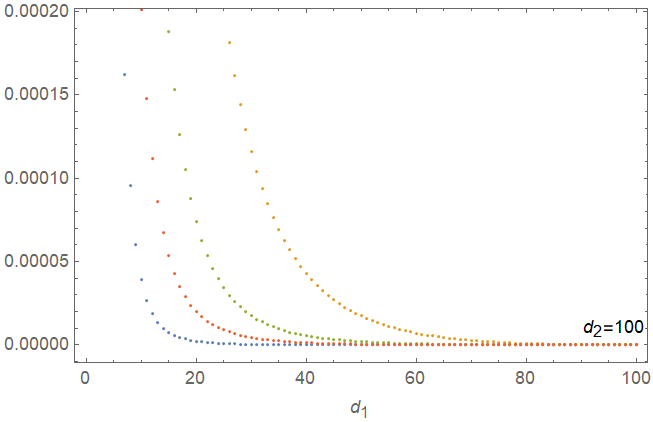}\caption{The difference of the thresholds which realize the equality in Eqs.
		\ref{eq:ineq_DV}-\ref{eq:ineq_ESIC} and \ref{eq:per} ($ p_{dV}-p_{ER } $ in blue, $ p_{E}-p_{ER} $ in red, $ p_{F}-p_{ER} $ in green, $ p_{R}-p_{ER} $ in orange) varying from $ d_1=2$ to $ d_2=100 $. This is an asymptotic pattern for high values of $ d_1, d_2 $. For low value of $ d_1,d_2 $ the pattern is not respected. In particular for low values of $d_{1}, d_2$
		$ p_E\leq p_{dV} $ and $ p_R\leq p_F $. For $ d_1=2,d_2=3 $ also $p_R\leq p_{dV}  $.}
	\label{fig:comparison}
\end{figure} 
	\begin{acknowledgments}
		DC and GSa were supported by the Polish National Science Centre project
		2018/30/A/ST2/00837. GSc is supported by Istituto Nazionale di Fisica
		Nucleare (INFN) through the project ``QUANTUM''. We acknowledge
		the Toru\'{n} Astrophysics/Physics Summer Program TAPS 2018 and the
		project PROM at the Nicolaus Copernicus University. GSc thanks S.
		Pascazio, P. Facchi and F. V. Pepe for invaluable human and scientific
		support, for suggestions and encouragements which led to the realization
		of the present work.
	\end{acknowledgments}
\appendix
\section{Proof of Theorem \ref{I}}
Introducing a vectorization of an operator \cite{Watrous,Gilchrist} $A=\sum_{i,j}A_{ij}|i\rangle\langle j|$
	via $|A\rangle=\sum_{i,j}A_{ij}|i\rangle\otimes|j\rangle$ one has
	$\mathcal{R}(A\otimes B)=|A\rangle\langle B^{*}|$, where the complex
	conjugation is taken w.r.t. the basis used for the vectorization.
	The resulting matrix $\mathcal{R}(\rho_{p})$ is the correlation tensor $C(\rho_{p})$
	for a choice of bases: $\{|e_{i}\rangle\langle e_{j}|\}_{i,j=1}^{d_{1}}\subset\mathcal{B}(\mathbb{C}^{d_{1}})$
	and $\{|f_{i}\rangle\langle f_{j}|\}_{i,j=1}^{d_{2}}\subset\mathcal{B}(\mathbb{C}^{d_{2}})$.
	These bases are orthonormal, but not hermitian, hence the matrix $C$
	can have complex entries, but its singular values and trace norm are
	the same as the case we choose hermitian orthonormal bases. One has
	\begin{align}
	  C = & \mathcal{R}(\rho_{p})=(1-p)\frac{1}{d_{1}d_{2}}\sum_{i=1}^{d_{1}}\sum_{j=1}^{d_{2}}|e_{i} \otimes e_i \rangle\langle f_{j}\otimes f_j |\nonumber \\
	& \qquad + \frac{p}{d_{1}} \sum_{i,j=1}^{d_{1}}|e_{i}\otimes e_j \rangle\langle f_{i}\otimes f_j | \\
	= & (1-p)\left|\frac{\oper_{\mathrm{d_{1}}}}{d_{1}}\right\rangle \left\langle \frac{\oper_{\mathrm{d_{2}}}}{d_{2}}\right| + \frac{p}{d_{1}} \sum_{i,j=1}^{d_{1}}|e_{i}\otimes e_j\rangle\langle f_{i}\otimes f_j| , \nonumber  \label{eq:Crho}
	\end{align}
where we used
$$   | \oper_{d_1}\rangle = \sum_{i=1}^{d_1} |e_i \otimes e_i \rangle \ , \ \  | \oper_{d_2}\rangle = \sum_{i=1}^{d_2} |f_i \otimes f_i \rangle . $$
To recast $D_{x}^{\left(1\right)},D_{y}^{\left(2\right)}$
	in the not hermitian basis we sandwich the above matrix with
$$  \mathbb{D}^{(1)}_x = \oper_{d_1} \otimes \oper_{d_1} +\frac{x-1}{d_{1}}|\oper_{\mathrm{d_{1}}} \rangle\langle\oper_{\mathrm{d_{1}}}| ,  $$
and
$$  \mathbb{D}^{(2)}_y = \oper_{d_2} \otimes \oper_{d_2} +\frac{y-1}{d_{2}}|\oper_{\mathrm{d_{2}}} \rangle\langle\oper_{\mathrm{d_{2}}}| .  $$
	In other words, $D_{x}^{\left(1\right)}$ is obtained replacing the first
	entry $1$ with $x$ using the projector $|\oper_{\mathrm{d_{1}}} \rangle\langle\oper_{\mathrm{d_{1}}}|/d_1$. Analogously for $D_{y}^{\left(2\right)}$. Then matrix $C_{xy}$ follows
\begin{widetext}
\begin{eqnarray}\label{}
  C_{xy} &=& \left(\oper_{d_1} \otimes \oper_{d_1} +\frac{x-1}{d_{1}}|\oper_{\mathrm{d_{1}}} \rangle\langle\oper_{\mathrm{d_{1}}}|\right) \Big( (1-p)\left|\frac{\oper_{\mathrm{d_{1}}}}{d_{1}}\right\rangle \left\langle \frac{\oper_{\mathrm{d_{2}}}}{d_{2}}\right| + \frac{p}{d_{1}} \sum_{i,j=1}^{d_{1}}|e_{i}\otimes e_j\rangle\langle f_{i}\otimes f_j| \Big) \left(\oper_{d_2} \otimes \oper_{d_2} +\frac{y-1}{d_{2}}|\oper_{\mathrm{d_{2}}} \rangle\langle\oper_{\mathrm{d_{2}}}|\right) \nonumber \\   &=&
  \frac{\left(y-p\right)x}{d_{1}d_{2}}\left|\oper_{\mathrm{d_{1}}}\right\rangle \left\langle \oper_{\mathrm{d_{2}}}\right| + \frac{p}{d_{1}}   \sum_{i,j=1}^{d_{1}}|e_{i}\otimes e_j\rangle\langle f_{i}\otimes f_j|
	+ p\frac{\left(x-1\right)}{d_{1}^{2}}   \sum_{i,j=1}^{d_{1}}|e_{i}\otimes e_i\rangle\langle f_{j}\otimes f_j|  .
\end{eqnarray}
\end{widetext}
Note that
\begin{eqnarray*} 
    \mathbb{D}^{(1)}_x C  \mathbb{D}^{(2)}_y =& U {D}^{(1)}_x U^\dagger C  V D^{(2)}_y V^\dagger  \\
 =&   U {D}^{(1)}_x C^{\rm can} D^{(2)}_y V^\dagger
\end{eqnarray*}
where $ C^{\rm can}=   U^\dagger C  V , $
defines the canonical correlation tensor. Hence, one has
\begin{equation}\label{}
  \|  C_{xy}  \|_1 = \| {D}^{(1)}_x C^{\rm can} D^{(2)}_y \|_1 . 
\end{equation}
To compute the trace-norm of $C_{xy}$ note that
	\begin{align}
	C_{xy}C_{xy}^{\dagger}= & \Bigg(\frac{x^{2}\left(y^{2}-p^{2}\right)}{d_{1}^{2}d_{2}}+\frac{p^{2}\left(x^{2}-1\right)}{d_{1}^{3}}\Bigg)\left|\bm{1}_{\mathrm{d_{1}}}\right\rangle \left\langle \bm{1}_{\mathrm{d_{1}}}\right|\nonumber \\
	& +\frac{p^{2}}{d_{1}^{2}}\oper_{d_{1}} \otimes \oper_{d_2} . 
	\end{align}
	Now, since $\left|\bm{1}_{\mathrm{d_{1}}}\right\rangle \left\langle \bm{1}_{\mathrm{d_{1}}}\right|$
	commutes with $\oper{d_{1}} \otimes \oper_{d_2}$, they share the same
	set of eigenvectors, therefore the spectrum $\sigma$ (expressed with
	the geometric multiplicity of the eigenvalues) reads
	\begin{align}
	\sigma\left(C_{xy}C_{xy}^{\dagger}\right)= & \left(d_{1}^{2}-1\right)\times\left\{ \frac{p^{2}}{d_{1}^{2}}\right\} \nonumber \\
	\cup & \Bigg\{\frac{x^{2}}{d_{1}d_{2}}\left(y^{2}+p^{2}\frac{d_{2}-d_{1}}{d_{1}}\right)\Bigg\}.
	\end{align}
Finally we have that separability of an isotropic state with unequal dimension $\rho_{p}$
	implies
\begin{align}
	\|C_{xy}\|_1 = & \frac{d_{1}^{2}-1}{d_{1}}p+\frac{x}{\sqrt{d_{1}d_{2}}}\sqrt{y^{2}+p^{2}\frac{d_{2}-d_{1}}{d_{1}}}\nonumber \\
	\le & \sqrt{\frac{d_{1}-1}{d_{1}}+\frac{x^{2}}{d_{1}}}\sqrt{\frac{d_{2}-1}{d_{2}}+\frac{y^{2}}{d_{2}}}.\label{Cxynorm}
\end{align}
In particular, for $x,y=0$, $x,y=1$, $x=\sqrt{2/d_{1}},y=\sqrt{2/d_{2}}$
	and $x=\sqrt{d_{1}+1},y=\sqrt{d_{2}+1}$, the above condition implies
	\begin{align}
	p\frac{d_{1}^{2}-1}{d_{1}}\le\sqrt{\frac{d_{1}-1}{d_{1}}}\sqrt{\frac{d_{2}-1}{d_{2}}},\label{eq:ineq_DV}\\
	\frac{d_{1}^{2}-1}{d_{1}}p+\frac{1}{\sqrt{d_{1}d_{2}}}\sqrt{1+p^{2}\left(\frac{d_{2}}{d_{1}}-1\right)}\le1,\\
	\frac{d_{2}\left(d_{1}^{2}-1\right)p+\sqrt{4+2p^{2}\left(\frac{d_{2}}{d_{1}}-1\right)d_{2}}}{\sqrt{d_{1}^{2}-d_{1}+2}\sqrt{d_{2}^{2}-d_{2}+2}}\le1,\\
	\frac{d_{1}^{2}-1}{d_{1}}p+\sqrt{\frac{d_{1}+1}{d_{1}d_{2}}}\sqrt{d_{2}+1+p^{2}\left(\frac{d_{2}}{d_{1}}-1\right)}\le2.\label{eq:ineq_ESIC}
	\end{align}
If the state $\rho_{p}$ is separable, then
	\begin{align}
	0 & \geq\frac{d_{1}^{2}-1}{d_{1}}p+\frac{x}{\sqrt{d_{1}d_{2}}}\sqrt{y^{2}+p^{2}\frac{d_{2}-d_{1}}{d_{1}}} -\mathcal{N}_{x,d_{1}}\mathcal{N}_{y,d_{2}}\label{eq:criterion_strong}
\end{align}
which obviously implies a weaker condition
\begin{align}
	0 & \geq\frac{d_{1}^{2}-1}{d_{1}}p-\mathcal{N}_{x,d_{1}}\mathcal{N}_{y,d_{2}}.\label{eq:criterion_weak}
	\end{align}
Let
	\begin{equation}
	p_{0}:=\frac{d_{1}}{d_{1}^{2}-1}\mathcal{N}_{x,d_{1}}\mathcal{N}_{y,d_{2}}.
	\end{equation}
It is clear that if $p>p_{0}$, then a weaker condition (\ref{eq:criterion_weak}) detects entanglement of $\rho_p$. If $p \leq p_0$, then  we can use a tighter inequality (\ref{eq:criterion_strong})
\begin{equation}
	-\frac{x}{\sqrt{d_{1}d_{2}}}\sqrt{y^{2}+p^{2}\frac{d_{2}-d_{1}}{d_{1}}}\geq\frac{d_{1}^{2}-1}{d_{1}}p-\mathcal{N}_{x,d_{1}}\mathcal{N}_{y,d_{2}}.
	\end{equation}
Under the assumption that we do not detect entanglement with the weaker
	condition \eqref{eq:criterion_weak}, both the terms of the inequality
	are negative, hence we can square and inverse the inequality:
	\begin{align}\label{eq:quadratic_inequality}
	  \mathcal{F}\left(p\right)
	  & = \left( \frac{d_{1}^{2}-1}{d_{1}}p-\mathcal{N}_{x,d_{1}}\mathcal{N}_{y,d_{2}} \right)^2 \nonumber \\
	  & - \frac {x^2}{d_{1}d_{2}} \left( y^{2}+p^{2}\frac{d_{2}-d_{1}}{d_{1}} \right) \nonumber \\
	  & = a_{d_{1},d_{2}}\left(x\right)p^{2}+b_{d_{1},d_{2}}\left(x,y\right)p+c_{d_{1},d_{2}}\left(x,y\right)\geq0,
	\end{align}
	where
	\begin{align}
	a\equiv a_{d_{1},d_{2}}\left(x\right)= & \frac{\left(d_{1}^{2}-1\right)^{2}}{d_{1}^2}-x^{2}\frac{d_{2}-d_{1}}{d_{1}^{2}d_{2}},\label{eq:a}\\
	b\equiv b_{d_{1},d_{2}}\left(x,y\right)= & -2\frac{d_{1}^{2}-1}{d_{1}}\mathcal{N}_{x,d_{1}}\mathcal{N}_{y,d_{2}},
	\end{align}
	\begin{align}
	c\equiv c_{d_{1},d_{2}}\left(x,y\right)= & \left(\mathcal{N}_{x,d_{1}}\mathcal{N}_{y,d_{2}}\right)^{2}-\frac{x^{2}y^{2}}{d_{1}d_{2}}.
	\end{align}
	A direct calculation shows always $\Delta=b^{2}-4ac\geq0$. Let $p_{\pm}=\frac{-b\pm\sqrt{\Delta}}{2a}$ {be the roots of $\mathcal{F}\left(p\right)$}
	.
	Notice, that $\mathcal{F}\left(p_0\right) \le 0$ (equality only for $x=0$), $c\geq0$, $b<0$. Hence Vieta's formulas implies
	\begin{enumerate}
		\item $a>0\Longrightarrow p_{xy}^{\left(+\right)}>p_{0}>p_{xy}^{\left(-\right)}>0$
		\item $a<0\Longrightarrow p_{0}>p_{xy}^{\left(-\right)}>0>p_{xy}^{\left(+\right)}$.
	\end{enumerate}
	In both cases the solution of the inequality (\ref{eq:quadratic_inequality}) reads as $p \in [0, p_-]$ and due to the continuity of $\mathcal{F}$ the limit formula for $a=0$ agrees with the solution of the linear inequality.
\section{Proof of Proposition 1}

The thresholds $p_E$ and $p_R$ are defined as the lowest roots of the quadratic polynomials:
	\begin{align}
	2(d_1^3d_2-2d_1d_2+1)p_R^2-4d_2(d_1^2-1)p_R\nonumber\\+2(d_1d_2-1)\stackrel{df}{=}f_R(p_R)=0 \label{fR} \\
	(d_1^3d_2-2d_1d_2+d_1-d_2+1)p_E^2-4d_2(d_1^2-1)p_E\nonumber\\+(3d_1d_2-d_1-d_2-1)
	\stackrel{df}{=}f_E(p_E)=0 \label{fE}
	\end{align}
	We will prove the relation between roots $p_E\le p_R$ by showing that all roots of $f_R$ and $f_E$ are in a set: $\{x: f_E(x) < f_R(x)\}$. To show it let us calculate the difference $f_E-f_R$:
	\begin{align}
	(f_E-f_R)(x) = -(d_1^3d_2-2d_1d_2+d_2-d_1+1)x^2 \nonumber\\
	+ (d_1-1)(d_2-1)
	\end{align}
	It is positive in the range $[-x_0,x_0]$, where:
	\begin{align}
	x_0=\sqrt{\frac{d_2-1}{(d_1+1)d_1d_2-(d_2+1)}}.
	\end{align}
	We will show that $f_R$ and $f_E$ are positive in the above range, showing that
	\begin{enumerate}
	 \item $f_R(x_0) = f_E(x_0)$,
	 \item $f_R$ is descending in $x_0$.
       \end{enumerate}
	One has
	\begin{align}
	f_R(x_0) =& f_E(x_0) \nonumber \\
	=& d_2\frac{2d_1^3d_2+d_1^2d_2-3d_1d_2-d_1^3-d_1^2+2}{(d_1+1)d_1d_2-(d_2+1)} \nonumber\\
	-&2d_2(d_1^2-1)\sqrt{\frac{d_2-1}{(d_1+1)d_1d_2-(d_2+1)}}.
	\end{align}
	Hence, we want to prove that
	\begin{align} \label{ineq_pEpR}
	2d_1^3d_2+d_1^2d_2-3d_1d_2-d_1^3-d_1^2+2 \nonumber\\
	- 2(d_1^2-1)\sqrt{(d_2-1)((d_1+1)d_1d_2-(d_2+1))} \ge 0.
	\end{align}
	To do this, we will rewrite the above as:
	\begin{align*}
	  (d_1^2-1) + d_1d_- \frac{2d_1+3}{2d_1+2} \ge \\
	  \sqrt{d_-^2(d_1^2+d_1-1)+d_-d_1(2d_1+3)(d_1-1)+(d_1^2-1)^2},
	\end{align*}
	where $d_-=d_2-d_1$. After squaring the latter simplifies to:
	\begin{displaymath}
	  d_1^2d_-^2(2d_1+3)^2 \ge 4d_-^2(d_1+1)^2(d_1^2+d_1-1),
	\end{displaymath}
	what finally gives:
	\begin{displaymath}
	  d_-^2(d_1^2+4d_1+4) = (d_2-d_1)^2(d_1+2)^2 \ge 0.
	\end{displaymath}
	We prove the second property showing, that the minimum of $f_R$ is greater than $x_0$:
	\begin{displaymath}
	  \frac{d_2(d_1^2-1)}{d_2d_1^3-2d_1d_2+1} > \sqrt{\frac{d_2-1}{d_1^2d_2+d_1d_2-d_2-1}}
	\end{displaymath}
	While $d_1 \le d_2$, we can estimate the RHS from above by $1/(d_1+1)$ and prove that the inequality holds for the estimation. The latter reduces to:
	\begin{displaymath}
	 0 < d_2(d_1^2-1)(d_1+1) - (d_2d_1^3-2d_1d_2+1) = d_2d_1^2 - d_2 - 1
	\end{displaymath}
	and holds for $d_1,d_2 \ge 2$.

	One has $f_E \ge f_R \ge 0$ in $[-x_0,x_0]$ and $f_E < f_R$ for $x> x_0$. The threshold $p_E$ and $p_R$ are roots on the left of the vertices of the parables $f_E$ and $f_R$ in Eq. \eqref{fR} and \eqref{fE} respectively. Both $p_E$ and $p_R$ are obviously greater than $x_0$ where $f_E < f_R$. It implies that $f_E$ reaches $0$ first and hence $p_E < p_R$.

\section{Proof of Theorem 3}

One finds
	\begin{align}
	\rho_{p}-\rho_{1}\otimes\rho_{2}&=\nonumber\\
	=\frac{p}{d_{1}}\sum_{i,j=1}^{d_{1}}&\left|e_{i}\right\rangle \left\langle e_{j}\right|\otimes\left|f_{i}\right\rangle \left\langle f_{j}\right| - \frac{p}{d_{1}^{2}}\oper_{d_{1}} \otimes \sum_{i=1}^{d_1} |f_i\rangle \langle f_i| ,
	\end{align}
and using again a vectorization technique one obtains 
	\begin{align}
	C_{\mathrm{ER}} \equiv &C\left(\rho_{p}-\rho_{1}\otimes\rho_{2}\right) \nonumber\\  
	=&\frac{p}{d_{1}} \sum_{i,j=1}^{d_1} |e_i \otimes e_j\rangle \langle f_i \otimes f_j|  - \frac{p}{d_{1}^{2}}\left|\oper_{\mathrm{d_{1}}}\right\rangle \left\langle \oper_{\mathrm{d_{1}}}\right|.
	\end{align}
$C_{\mathrm{ER}}$ stands for enhanced realignment correlation matrix
	and the spectrum of $C_{\mathrm{ER}}$$C_{\mathrm{ER}}^{\dagger}$
	reads
	\begin{equation}
	\sigma\left(C_{\mathrm{ER}}C_{\mathrm{ER}}^{\dagger}\right) = \left\{ \frac{p^{2}}{d_{1}^{2}}\right\} \times\left(d_{1}^{2}-1\right)\cup\left\{ 0\right\}.
	\end{equation}
	This brings to the condition
	\begin{equation}
	\frac{d_{1}^{2}-1}{d_{1}}p \leq \sqrt{ \frac{d_{1}-1}{d_{1}}}\sqrt{\frac{d_{2}-1}{d_{2}}-p^{2}\left(\frac{d_{2}-d_{1}}{d_{1}d_{2}}\right) }.
	\end{equation}
	The equality holds for
	\begin{equation}
	p_{\mathrm{ER}}=\sqrt{ \frac{d_{2}-1}{d_{2}\left(d_{1}^{2}+d_{1}-1 \right)-1} } .
	\end{equation}

\bibliographystyle{unsrt}

\end{document}